\documentclass[prb,twocolumn,nofootinbib]{revtex4-1}
\usepackage{graphicx}
\usepackage{bm}
\usepackage{color}
\usepackage{braket}
\usepackage{amsmath}
\usepackage{cleveref}

\begin{document}
\title{Hybrid gausslet/Gaussian basis sets}
\author{Yiheng Qiu}
\affiliation{Department of Physics and Astronomy, University of California, Irvine, CA 92697-4575 USA}

\author{Steven R. White}
\affiliation{Department of Physics and Astronomy, University of California, Irvine, CA 92697-4575 USA}
\date{\today}

\begin{abstract}
We introduce hybrid gausslet/Gaussian basis sets, where a standard Gaussian basis is added to a gausslet basis in order to increase accuracy near the nuclei while keeping the spacing of the grid of gausslets relatively large.  The Gaussians are orthogonalized to the gausslets, which are already orthonormal, and approximations are introduced to maintain the diagonal property of the two electron part of the Hamiltonian, so that it continues to scale as the second power of the number of basis functions, rather than the fourth. We introduce several corrections to the Hamiltonian designed to enforce certain exact properties, such as the values of certain two-electron integrals. We also introduce
a simple universal energy correction which compensates for the incompleteness of the basis stemming from the electron-electron cusps, based on the measured double occupancy of each basis function.   We perform a number of Hartree Fock and full configuration interaction (full-CI) test calculations on two electron systems, 
and Hartree Fock on a ten-atom hydrogen chain, to benchmark these techniques.  The inclusion of the cusp correction allows us to obtain complete basis set full-CI results, for the two electron cases, at the level of several  microHartrees, and we see similar apparent accuracy for Hartree Fock on the ten-atom hydrogen chain.

\end{abstract}
\maketitle

\section{Introduction}
The introduction of Gaussian-type basis sets into quantum chemistry by Boys, Pople
and others \cite{Boys50,Pople71,Clementi65,Whitten66,Davidson86}, decades ago represented a major advance,
and these basis sets are still by far the most widely used in chemistry. 
A strength of Gaussian bases is their compactness; one generally needs many fewer Gaussians
than, say, plane waves, for comparable systems, since the completeness of plane waves is uniform and general-purpose,
rather than adapted not only to the nuclear cusp but to the specific properties of
each type of atom.
However, Gaussians also have weaknesses, which are becoming more noticeable as
new many-electron correlation algorithms have been developed. One weakness is the $N^4$ scaling of the 
number of two-electron
integrals with the basis size $N$, which is particularly inconvenient for density matrix renormalization
group (DMRG) calculations, where techniques to improve the scaling, such as density fitting, 
have not proven useful. 
The large number of Hamiltonian terms also is problematic for quantum computing, where the exponential
size of the many-particle Hilbert space is not a problem, but the
interaction terms in the Hamiltonian must be implemented with quantum gate operations.
Gaussians are also non-orthogonal, and orthogonalization procedures tend to make tails which
decay slowly in distance, particularly when the basis is being made more complete.  This non-locality of the basis limits one's ability to make the Hamiltonian more sparse by throwing away longer-ranged terms. 

The convergence of DMRG depends on the entanglement across the system. The entanglement
can be minimized at the one particle level by rotating to molecular orbitals (``energy localization''), but
with interactions the use of molecular orbitals leads to volume law entanglement, dominating the area law
entanglement of a local basis. (This has been shown explicitly in the Hubbard model \cite{ESLN05}.) Probably the best approach is to energy-localize some of the degrees of freedom (i.e. the highly-occupied orbitals) while spatially localizing nearly degenerate valence degrees of freedom\cite{OHNSYC15}. Optimizing the choice of basis for the most efficient
DMRG calculations, taking into account both entanglement and Hamiltonian complexity,  is an area of active study \cite{KVLE16}. With standard Gaussian bases, one has difficulty in exploring  optimal localization in the very localized regime, due to the orthogonalization tails. 

With these issues in mind, we have been developing gausslet basis sets. Gausslet bases allow the
constructions of accurate Hamiltonians with only $2N^2$ terms, an $N\times N$ matrix for the interactions,
and another for the single particle terms, and are highly local. Gausslets live on a distorted grid,
and their completeness can be improved systematically and smoothly by decreasing the spacing between
gausslets.  On the other hand, gausslet
bases can be much bigger than Gaussian bases.  
If the gausslets are used as an intermediate basis allowing convenient construction of a much smaller
basis, as in the gausslet discontinous Galerkin method\cite{McClean20}, then the size of the gausslet basis is not a 
problem.
For use directly in, say, a DMRG calculation, one would strongly prefer that the basis size not
be too much larger than Gaussian bases. This requires careful construction.
In addition, electronic structure algorithms should be adapted to make use of the 
reduced scaling of Hamiltonians granted by gausslets.

In the first use of gausslets for realistic electronic structure calculations\cite{White19}, 
the combination
of gausslets and DMRG was demonstrated to produce more accurate complete basis set, full-CI energies for
ten equally spaced hydrogen atoms in a chain, than any method (including DMRG) using Gaussian bases.
However, the pure gausslet bases used there would be difficult to apply beyond hydrogen or helium
systems, at least without the use of pseudopotentials to deal with the core degrees of freedom. 
Gausslets by themselves become inefficient if they must represent the huge range of length scales
present with large $Z$ atoms.
Here we develop hybrid Gaussian/gausslet bases, where the Gaussians are utilized to efficiently
represent the core region. We show how, despite the presence of Gaussians, a purely diagonal $N^2$ interaction
Hamiltonian can be formed with high accuracy.

In the next section, we give an overview of gausslet bases. 
In Section III, we explain how we add Gaussian bases to the gausslets, discussing orthogonalization, contraction, and showing the effects of Gaussian on the one-body part of Hamiltonian. 
In Section IV, we discuss several diagonal approximations for the two-electron integrals of the added Gaussians. Numerical result are shown comparing these different approximations.
In Section V, we present various techniques making nuclear and interaction corrections, by renormalizing the one and two-electron integrals. 
In Section VI, we study the scaling behaviors of energy errors and double occupancies. Therein, a correction is derived accounting for the inaccuracy in describing the electron-electron cusp. 
Finally, in Section VII, we present a summary and conclusions.

\section{Gausslets}
A one dimensional Gaussian is a minimum uncertainty function, with simultaneous locality
both in real and Fourier space, a highly desirable property for a basis function. 
A grid of evenly spaced Gaussians, with width about equal to the spacing, has excellent
completeness properties, but the basis is not orthogonal.  Orthogonalizing it produces functions
with long tails, spoiling the spatial locality. In fact, constructing 1D orthogonal bases with 
locality in both real and Fourier space was a difficult and long-standing problem, which 
was only solved in a satisfactory way by the development of wavelets.

\begin{figure}[t]
    \includegraphics[width=\linewidth]{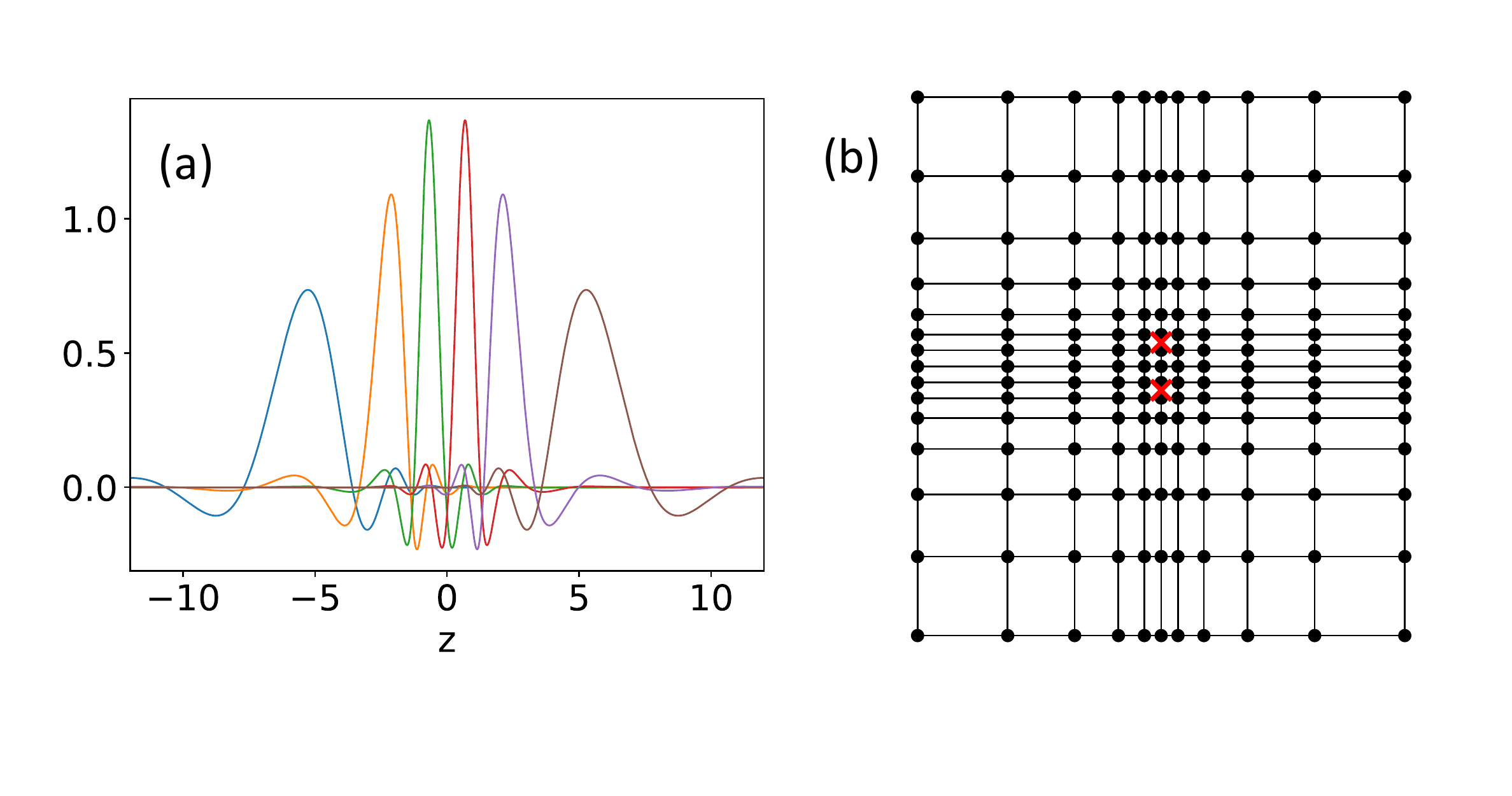}
    \caption{(a) Distorted gausslets along the $z$ direction for the H$_2$ molecule, with nuclei at $z=\pm 2$. Only every other gausslet is shown for clarity. 
    (b) The distorted grid showing the centers of the gausslets in the $xz$ plane. The red crosses denote the positions of 
    the two hydrogen nuclei.
    \label{Fig:gausslet_1d_and_grid}} 
\end{figure}

A gausslet is a new kind of basis function based on wavelet theory, as well as non-standard techniques borrowed
from tensor networks and quantum circuit theory\cite{White17,Evenbly18}.
An example of one-dimensional gausslets is shown in Fig. \ref{Fig:gausslet_1d_and_grid}(a).
The non-standard techniques give gausslets several very desirable properties for use as
a basis for quantum problems. Gausslets  are exactly orthogonal.
They are complete, able to represent any polynomial up to a specified order to near floating-point
accuracy. (Successful
construction was only accomplished up to order 10, and that is the order gausslets we use here.)
They are highly localized, with exponential decay out to a certain range, followed by much
faster $e^{-\alpha x^2}$ Gaussian decay at longer distances, making them for practical purposes
compact. They have similar behaviors in Fourier space. It is impossible to get true compactness
simultaneously in both real and Fourier space, but gausslets, like Gaussians, mimic it closely.
Conventional compact wavelet scaling functions, such as the famous functions of Daubechies, 
due to their construction as a fixed point, and their perfect compactness, have only exponential decay in Fourier space, making
them hard to integrate numerically to high accuracy without special tricks. Gausslets,
even in the coordinate-transformed form used here, are very easy to integrate numerically to very
high accuracy.  (In the absence of the coordinate transformations, their integrals can be performed analytically.)

Gausslets are exactly symmetric. Conventional compact wavelet scaling functions are not, because
they arise from a factor-of-two scaling transformation.  Gausslets come from a factor of three
scaling transformation, allowing exact symmetry.\cite{White17,Evenbly18}  Gausslets have a compact form
as a sum over a uniform set of Gaussians, living on a grid one third the spacing of the gausslets
themselves.  Thus given the contraction coefficients, a gausslet is exactly defined
and ready for use.  Finally, gausslets have a very important moment property which makes them
integrate with smooth function just like $\delta$-functions.  Specifically, if
 $G(x)$ is a gausslet centered at the origin, then 
\begin{equation}
    \int G(x) x^m \mathrm{d}x = 0,
\end{equation}
for all integers $m$ ranging from $1$ to twice the completeness order of the gausslet, so
typically up to about $m=20$.  
An immediate implication of the moment property is that 
\begin{equation}
    \int G(x-x_0) f(x) \mathrm{d}x \approx f(x_0)/w,
\end{equation}
where $f(x)$ is any function that is smooth enough 
(i.e. high order derivatives are both continuous and small in magnitude), and 
$w=\int G(x) \mathrm{d}x$ is the self integral of the gausslet.
Consider, for simplicity, a single-particle one dimensional Schr\"odinger equation. The $\delta$-function property means that a potential term $V(x) G(x-x_0)$ can be replaced to high accuracy with $V(x_0)G(x-x_0)$.  Due to the orthogonality of the gausslets, this means that the potential energy matrix for $V$ can be written in
diagonal form, i.e. $V_{ij} = \delta_{ij} V(x_i)$.
Since 3D gausslets are products of the 1d gausslets, the same property holds in 3D. More importantly, the two-electron interaction terms also have an 
accurate diagonal form, $V_{ijkl} \to \delta_{ij} \delta_{kl} V_{ik}$.  In practice, we generally do not use the diagonal approximation for the one electron
potential, since its computational cost is small; we only use it for the two-electron interaction.

To represent ground-state wavefunctions more efficiently, a higher resolution around
the nuclei is needed. 
Thus, it is desirable to distort the 3D lattice so that more gausslets are placed near the nuclei,  
as shown in Fig. \ref{Fig:gausslet_1d_and_grid}(b). 
The distortion is implemented through a coordinate mapping along the three dimensions.
For example, the x-direction gausslet $G(x)$ is replaced by $\sqrt{u'(x)}G(u(x))$, 
where $u(x)$ is designed to increase the density of gausslets near the $x$-coordinates of the  nuclei.
For atomic systems, $u(x)$ is taken to be\cite{White17,White19}
\begin{equation}
    u(x) = \frac{1}{s}\sinh^{-1}(\frac{x}{c})
\end{equation}
where $x$ represents the distance from the nucleus, the spacing parameter $s$ determines the overall density of gausslets. The spacing of the gausslets decreases as one moves from the tails to the nucleus, but levels off near the core, as controlled by the core-size parameter $c$.  The gausslet spacing at the nucleus is $sc$. A distorted 3D gausslet is simply a direct product of distorted 1D gausslets along the three directions.
For molecular systems, the coordinate transformation is taken to be
the sum of the atomic ones, with a slight modification. Since the density of basis functions of the atoms overlap, and a simple sum would give a steadily decreased spacing as more atoms were added, we make each $c$ for each atom (and each coordinate direction) adjustable parameters, and adjust them so that the spacing at each nucleus in each direction is always exactly $sc$. (For a linear chain, only the $c$'s along the chain direction need adjusting.)
An example of distorted gausslets along the bond direction of H$_2$ molecule is shown in Fig. \ref{Fig:gausslet_1d_and_grid}(a).

This form of distorting the 3D gausslets is simpler than the multisliced scheme used in White and Stoudenmire\cite{White19}. To distinguish the two, we will call the simpler scheme coordinate-slicing.
In both schemes, each 3D gausslet is a product of 1D gausslets, but
in multislicing, different distortions for the $y$ gausslets are used for each value of $z$, and different distortions of the $x$ gausslets are used for each combination of $z$ and $y$. In coordinate slicing, the same $x$-distortion is applied to all $x$ gausslets, and the same for $y$ and $z$.  
The multisliced scheme reduces the number of gausslets needed, but integral evaluation (and programming complexity) is reduced in coordinate slicing.  More importantly, 
coordinate slicing leaves us with basis function on a (distorted) rectangular grid whereas multislicing does not.  The rectangular grid allows for further decimation using wavelet techniques which we will describe in a subsequent paper. This type of decimation can reduce the number of sites beyond that of multislicing. To reiterate, in this paper we utilize only coordinate slicing,
which is simpler, but results in a larger basis than cleverer schemes.

While undistorted gausslets have analytic integrals, distorted gausslets do not, but their very high degree of smoothness allows very accurate numerical integrals, and their product form over the three coordinate directions reduced the dimensionality of the integrals. The  Gaussian basis functions (see below) also have the same properties, and were numerically integrated in the same way,  as described in Ref. \onlinecite{White19}.

\section{Combining Gausslets with Gaussians}
The complete basis set (CBS) limit can be approached smoothly by decreasing the spacing between gausslets,
but the convergence can be slow due to the presence of the nuclear cusps.
A large number of gausslets are needed in regions close to nuclei in order to represent the nuclear cusps accurately. Gausslets are inefficient in this region for two reasons. First,  the simple 1D form of the distortion is less efficient than a true 3D distortion, which we do not do because the integrals defining the Hamiltonian would be much more difficult. Second, the distortion cannot be made too sharp---the scale cannot be changed too rapidly---or the completeness and moment properties of the gausslets will be damaged. 
On the other hand,
Gaussian orbitals are tailored to describe the nuclear cusps using a small number of functions per  atom.
In this sense, traditional Gaussians basis functions, such as cc-pVnZ, are complementary to gausslets.
Combining the Gaussians with gausslets eliminates the need of a large number of gausslets near the core regions. However, the higher angular momentum components from the Gaussians are not needed---these degrees of freedom are more efficiently represented by the gausslets, so we set a maximum angular momentum number $l$ and only add in Gaussians with the lower angular momentum . 

Note that the ability to add in core functions is due to the gausslets being a basis set, rather than a finite-difference grid.  We do not know how to combine basis functions with non-basis grids.  

Since the Gaussian orbitals are not orthogonal to the gausslets, 
we orthogonalize them to the gausslets and against each other.
We call these orthonormalized Gaussians  {\it residual Gaussians} here, 
and it is the residual Gaussians that are added into the basis set, rather than the original Gaussians.
However, the single particle integrals are done with the nonorthogonal Gaussians and gausslets, rather than the residual Gaussians, and an overlap matrix is used to transform the matrices into the orthonormal basis.

\begin{figure}[t]
    \includegraphics[width=\linewidth]{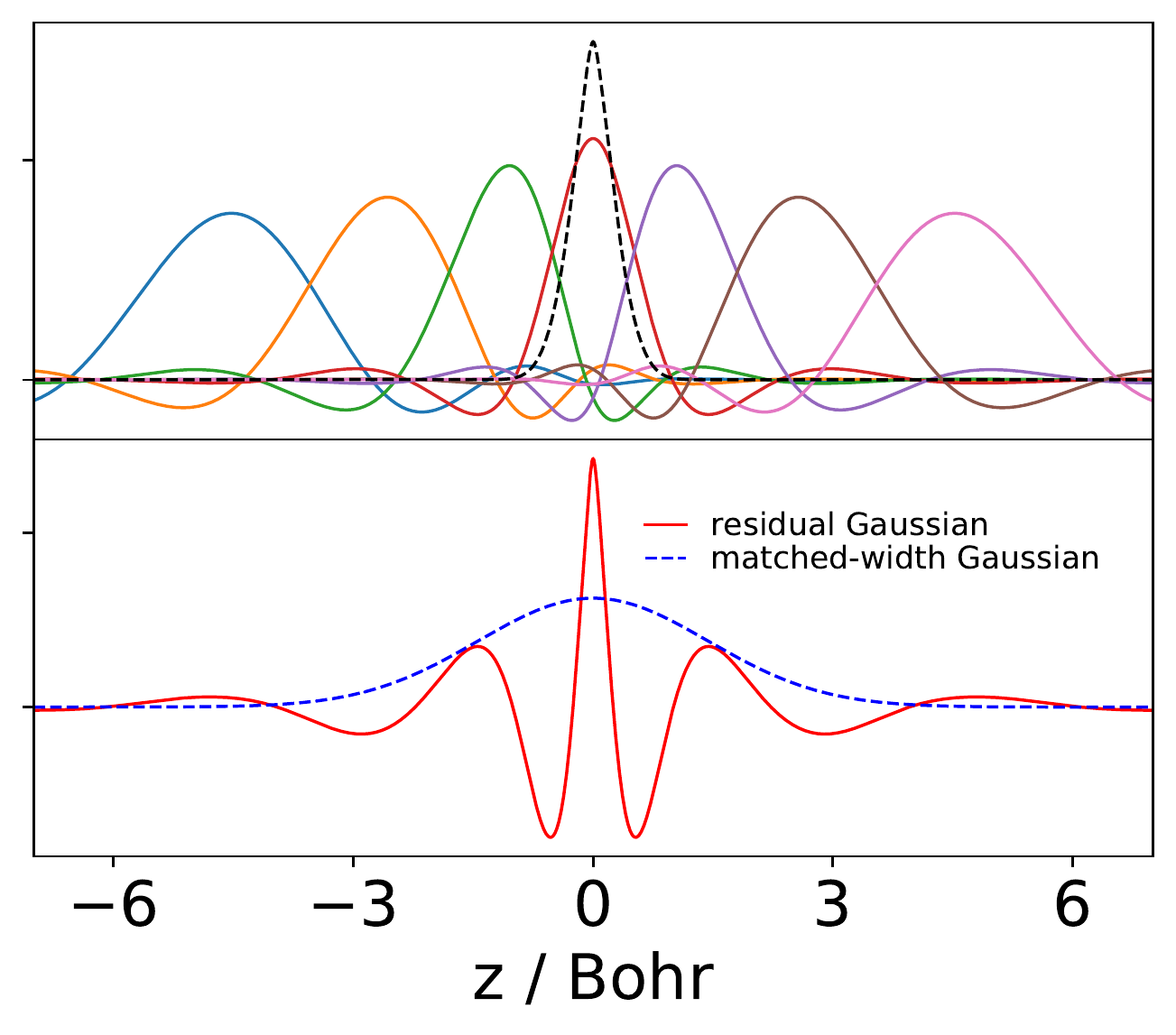}
    \caption{
    A 1D residual Gaussian, giving an easy-to-visualize illustration of the 3D residual Gaussians in the actual basis sets.
    Solid lines in the upper panel are distorted gausslets (with $s=c=1.0$), which are orthonormal, 
    while the dashed line in the upper panel is a contracted Gaussian, approximating a hydrogen 1s Slater orbital located at the origin, taken from the cc-pV6Z basis set.
    The solid line in the lower panel shows the resulting residual Gaussian after the contracted Gaussian is orthogonalized to the gausslets.
    The dashed line in the lower panel shows a Gaussian with a width that matches that of the residual Gaussian.  This Gaussian is used in approximating two electron integrals for the residual Gaussians.
    \label{Fig:res_gau}}
\end{figure}

To illustrate the features of residual Gaussians, an example of a 1D hybrid gausslet/Gaussian basis set is shown in Fig. \ref{Fig:res_gau}.
The system consists of one (1D) ``hydrogen" atom at the origin.
The residual Gaussians add just the necessary supplements to the gausslets to reproduce the nucleus cusp.
Note that the overall shape of the residual Gaussian is similar to that of a wavelet, since it is living in the high-frequency space,
but the residual Gaussian has a much sharper cusp at the core, inherited from the contracted Gaussian.

\begin{figure}[t]
    \includegraphics[width=7cm]{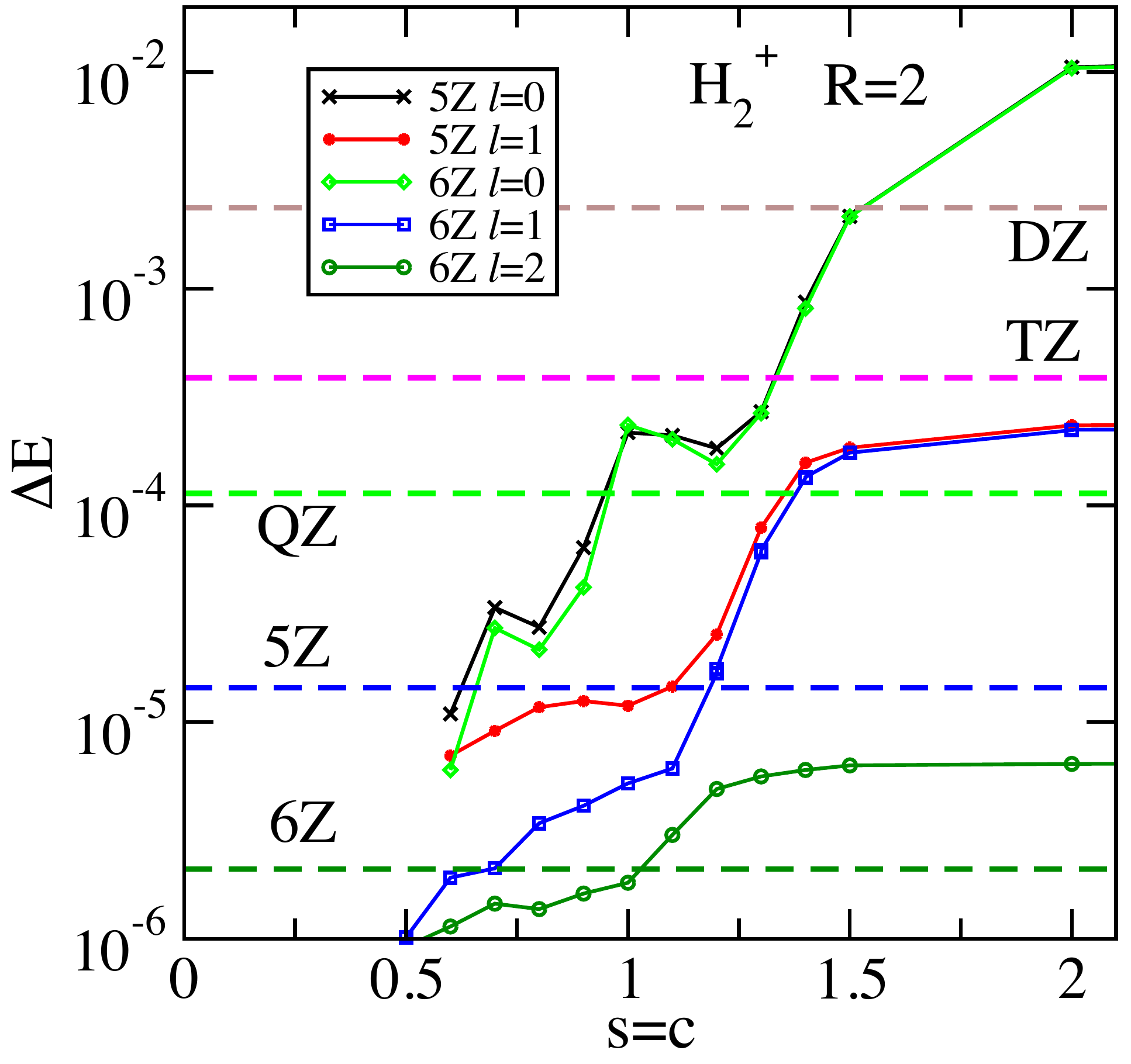}
    \caption{
    Assessing the effect of adding different Gaussian orbitals on the one-body Hamiltonian.
    Errors of H$_2^+$ ground-state energies (in a.u.) for pure Gaussian and for hybrid gausslet/Gaussian bases are shown, versus the spacing parameters $s$ and $c$ (which are equal here) of the gausslets. The horizontal dashed lines show energies of the pure Gaussian bases cc-pvNZ for N=D, T, Q, 5, and 6. The curves with symbols are the hybrid basis sets, labeled both by the Gaussian basis set and the maximum angular momentum $l$ used from the Gaussians (e.g. for $l=0$, only $S$ Gaussians were used).
\label{Fig:H2+}}
\end{figure}

To demonstrate the effect of adding Gaussians to gausslets at the 3D one-particle level, the ground-state energies of $H_2^+$ are shown in Fig. \ref{Fig:H2+}. For large $s,c$, the gausslet part of the basis does not contribute much to the completeness, but the curves do not tend to the pure Gaussian lines because the hybrid bases are missing the higher angular components of the Gaussians. For more moderate $s,c$ the  hybrid basis matches and passes the accuracy of the corresponding full Gaussian basis.  The sweet-spot we use in our work corresponds to $0.5 \le s,c \le 1.0$, with $l=1$ or $l=2$, using 5Z or 6Z basis sets.  Using only the $S$ Gaussians would limit accuracy to near $10^{-4}$ without much benefit in reduced basis size.  

\begin{figure}[t]
    \includegraphics[width=7cm]{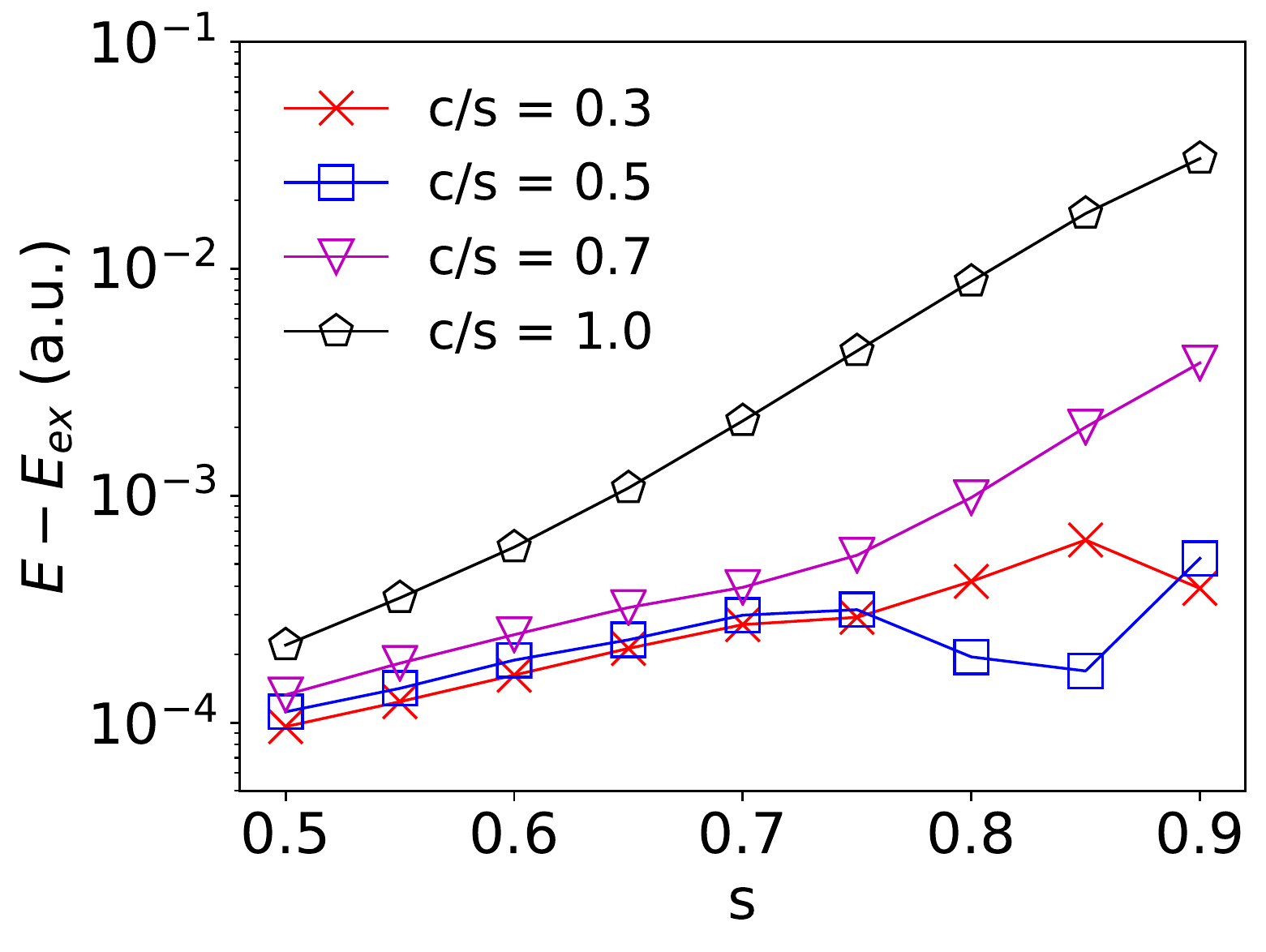}
    \caption{
    For different ratios of $c/s$, errors in the 
    ground state energy of the He atom are compared to the exact value ($-2.903724\ldots$), versus $s$.
    The S and P functions of the cc-pv5Z basss were added to the gausslets.  The two-electron integrals of the residual Gaussians are obtained through the MWG approximation (Section V).
\label{Fig:corefac_He}}
\end{figure}

The accuracy and size of a gausslet basis set is controlled primarily by the two transformation parameters $s$ and $c$.
For hydrogen systems, we find that $c=s$ is a good choice balancing accuracy and cost.
For larger $Z$ atoms, note that the width of the 1S orbital
 is proportional to $1/Z$, suggesting $c=s/Z$ could be a good choice 
 for generic atoms.
In Fig. \ref{Fig:corefac_He}, different ratios of $c/s$ are compared for a helium
atom. 
Significant improvement can be obtained by dropping the ratio $c/s$ from $1.0$ down to $0.5$, 
but little further improvement is gained by going from $0.5$ to $0.3$.
Similar calculations on a beryllium atom (not shown) are consistent with  $c/s = 1/Z$ being a  good choice.

For some systems, Gaussian bases can become linearly dependent, where the overlap matrix gets one or more eigenvalues close to zero.  In a basis with Gaussians only, one approach to dealing with this issue is to throw out the near-zero eigenvectors of the overlap matrix from the basis, reducing the size of the basis. In a gausslet/Gaussian basis, similar behavior can occur in the overlap matrix of the residual Gaussians after they have been orthogonalized to the gausslets.  In this context, throwing out eigenvectors may increase the spatial extent of the remaining functions, which we try hard to avoid.  We have only considered this problem in the context of chain systems.  In hydrogen chain systems one encounters linear dependence for longer chains at smaller values of the spacing $R$.  More complete bases makes the problem worse.

Our solution is to treat the larger width Gaussians separately, and to reconstruct them using gausslets as a sliced basis.  We assume a chain with the atoms along the $z$ axis. Linear dependence is generally caused by  Gaussians which are wide compared to the spacing $R$ of the chain. (This is clearly seen with an array of 1D Gaussians.) We remove from the basis the larger width Gaussians, with width $w>R/2$, where the width in terms of the Gaussian exponent is given by $w=1/\sqrt{2\zeta}$. For each $\zeta$, we replace the corresponding set of 3D Gaussians by (for S functions) $G^z_i(z) \exp(-\zeta (x^2+y^2))$, where the
$G^z_i(z)$ are the $z$-direction 1D distorted gausslets used in constructing the gausslet part of the basis. For each transverse Gaussian, this mixed set of functions is orthogonal. The cutoff we use for replacing the Gaussians, $R/2$, is a reasonable choice which has worked well.  This replacement is justified because the wide $z$-Gaussians are easily represented by the 1D $z$ gausslets.  

\section{Diagonal Approximation for Residual Gaussians}
The two-electron integrals involving the residual Gaussians do not have the diagonal property.
However, since the gausslets have reasonable completeness even near the cores and the residual Gaussians are orthogonal to them, the occupancy of the residual Gaussians is quite low, say around $\sim 10^{-4}$ for the systems studied here.  Therefore, we can make relatively crude approximations for their two-electron integrals without significantly affecting our results. We want to construct an accurate approximation for the two electron integrals involving any residual Gaussians, which is of diagonal form, like the gausslet-only terms, and which is not too costly to compute. 
We have considered a few simple approximations. In general, in order to keep the residual Gaussians as localized as possible,   we apply symmetric orthogonalization using the inverse square root of the overlap matrix. The residual Gaussians are highly oscillatory, since they are orthogonal to the gausslets. Consider the product of a residual Gaussian $g(\vec r)$ and another basis function $h(\vec r)$,  $gh$. This product integrates to zero (orthogonality) and is composed almost entirely of high-momentum components.  When viewed as a charge density, the potential of $gh$ will be very local.  This makes the diagonal approximation, where all two-electron integrals involving $gh$ are neglected, arise naturally. There would still be off-diagonal terms where all four indices are close together, which we neglect, due to the low occupancy of $g$.   Then, we need to specify how we calculate the diagonal terms, $V_{ij}$.  

In the first approximation, we
simply transfer the two electron integrals from the gausslet closest to the center of each residual Gaussian. We call this the Gaussian-gausslet transfer approximation (GGT).  The primary virtues of this approach are simplicity and low computational cost. Perhaps surprisingly, GGT gives generally satisfactory results.
Its primary weakness appears to be that some residual Gaussians may be substantially different in size than the nearby gausslet. 

This flaw is  addressed in our second approximation.
To avoid the difficult computational task of evaluating two-electron integrals directly with residual Gaussians, we construct an effective Gaussian orbital $G_{\rm eff} = G_x(x) G_y(y) G_z(z)$, where $G_x(x) = N \exp(-\gamma_x (x-a_x)^2)$, etc, and where $N$ is the normalization factor. We adjust the widths and centers to match  those of the  residual Gaussians, which are calculated exactly.
To obtain the widths and centers, we calculate the matrix $X$ representing the overlaps $\langle i | x | j\rangle$, and similarly for $x^2$, $y$, $y^2$, $z$, and $z^2$.  Then it is not costly to transform these matrices with the orthogonalizing matrices to obtain the corresponding values for the residual Gaussians.  
Typically, the residual Gaussians are wider than the pure Gaussians they came from. 
Then we evaluate all diagonal two electron integrals involving the residual Gaussians 
using these effective Gaussian orbitals.
We call this the matched-width-Gaussian (MWG) approximation.  Note that constructing and transforming the two electron integrals without this sort of approximation would be much too costly, since we would need many integrals in the non-diagonal $V_{ijkl}$ form to construct the diagonal terms we want.

Because a good diagonal approximation is so useful, we have studied a third, more involved way to approximate the two-electron integrals of residual Gaussians using wavelet theory. This approach involves a larger intermediate basis that contains both the gausslets and wavelets, but reduces the basis back to the gausslet size by contracting all wavelets into the gausslets.  This approach is discussed in the Appendix.  Despite being more complex and time-consuming than the MWG method, it did not give clear advantages in terms of accuracy, and we generally prefer the MWG method.

\begin{figure}[t]
    \includegraphics[width=\columnwidth]{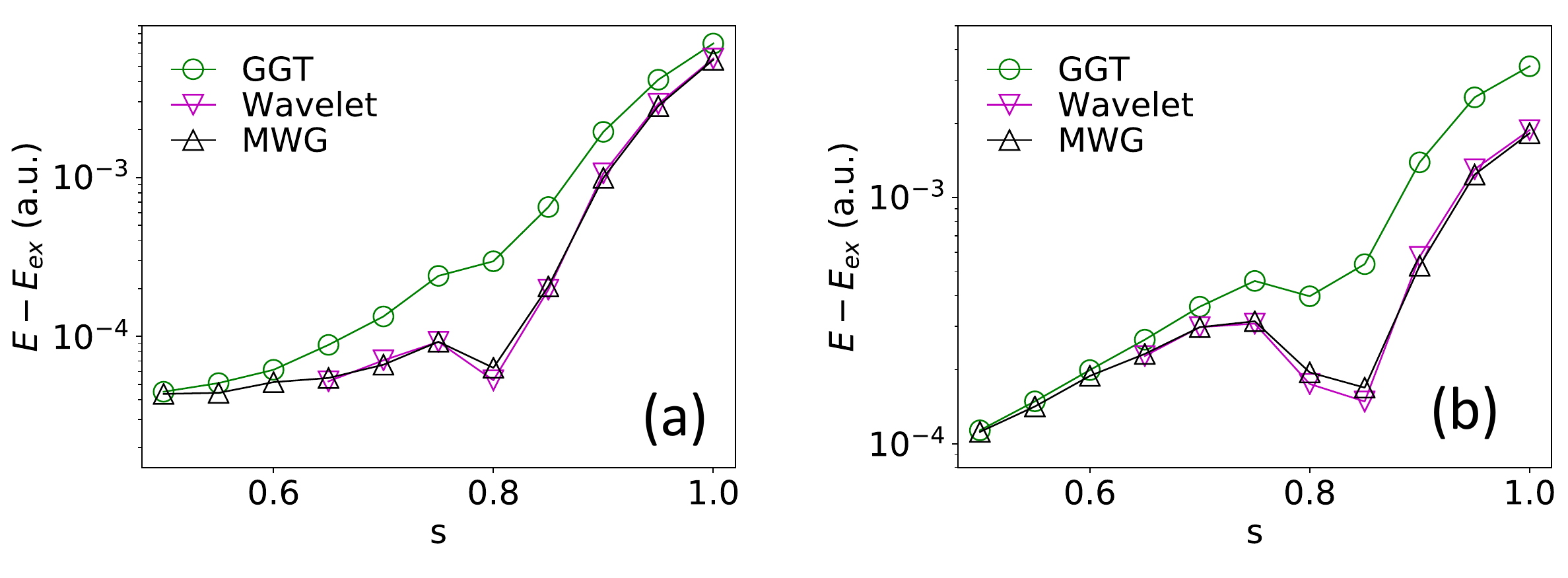}
    \caption{
    Comparing different diagonal approximations of the two-electron integrals involving residual Gaussians.
    Basis set induced errors in the energy of the He atom within (a) Hartree-Fock and (b) exact diagonalization are shown,
    versus $s$ with $c=0.5s$.
    The cc-pV6Z Gaussians, with up to P orbitals, are added.
    Three different approximations of the residual Gaussian two-electron integrals are shown. The very simple GGT approach is reasonable; the other two more sophisticated approaches perform almost identically.  
\label{Fig:He_wave}}
\end{figure}

Fig. \ref{Fig:He_wave} summarizes the accuracy of these diagonal approximations for the residual-Gaussian two-electron integrals, applied to the He atom either within the Hartree-Fock approximation or with an exact diagonalization. 
The three curves in each panel differ only in the way the diagonal two electron terms were obtained.
All three approximations achieve roughly similar accuracy, with the simple GGT approach somewhat less accurate.   The close similarity between the
results reflects both the small occupancy of the residual Gaussians and the success of the diagonal approximations. In the rest of the paper, we will use the MWG approximation.

\section{Correcting the Hamiltonian}
The replacement of the standard $V_{ijkl}$ two electron interaction with a diagonal approximation is an example of a more general approach. As we have presented it, its key property is that it does not interfere with  the rapid convergence of the energies and wavefunctions  to the precise complete basis set limit as the basis scale parameters are reduced. We can take a small step further and also consider {\it corrections} to the Hamiltonian that accelerate the convergence to the complete basis limit, leaving the final complete basis set answer unchanged and exact.  In principle, any correction which does not change
the limiting result and which gives better results with smaller basis size, or makes the convergence more systematic, is potentially useful.

The simplest types of corrections to our Hamiltonians are related to the behavior of the core.  
These corrections might be  important since the energy scales are large in the core region, and the diagonal approximations may induce noticeable errors there. Note there are no single-particle corrections needed---this is taken care of by the addition of the Gaussians, and we keep the full single particle matrix with no diagonal approximation. Thus we consider corrections which try to reduce errors associated with the two-particle diagonal approximation in the core region, where gausslets and Gaussians are both important. In the end, the corrections we describe here have only a small effect, which we interpret as an indication that  MWG for approximating the two-electron terms does a good job. (Later on, we will describe an electron-electron cusp energy correction which has a much bigger effect.) Further corrections to the two electron terms might be made if one went beyond the diagonal approximation in some limited way, such as incorporating terms involving three basis functions.  
We  describe three corrections in this section, which are abbreviated as the ``Exact Slater Orbital Integrals" (ESOI),  ``Exact Gaussian Orbital Integrals" (EGOI), and ``Stationary Fock" corrections. The Stationary Fock corrections should only be done in combination with EGOI---so after this section we do not consider them separately. 

{\it \underbar{Exact Orbital Integral corrections}}
Certain important orbital functions have exactly known two-electron integrals.  
A particular key case is the hydrogen 1s orbital
for which the Coulomb self integral $\bra{1s\,1s}V\ket{1s\,1s}=\frac{5}{8}Z$, where $Z$ is the atomic number of the nucleus.
Another important case is the Gaussian orbitals, where the two electron integrals are also analytically available.  The idea of this approach is to fix the Hamiltonian to satisfy the known integrals, but then we hope that this same correction helps more generally.  For example, suppose, hypothetically, that the diagonal approximation underestimated (or overestimated) interactions near the nuclei, because of the nuclear cusp. Then if a well-designed correction fixed the two-electron integral for one 1S approximate HF orbital by strengthening the interactions near the core, it would probably also mostly fix
an exact 1s HF orbital or a 1s natural orbital. 

Suppose the two-electron integrals of a set of target orbitals $\{\phi_I\}$  are known. 
If the hybrid basis set spans these orbitals, they can be expanded as a linear combination of the basis functions with coefficients $\phi_{iI}$. (For the case of Gaussian target orbitals, these are taken from the set of Gaussians we have already added to make the hybrid basis, so the basis does span the targets.)
In the diagonal approximation, our basis gives for the set two electron integrals of the target orbitals
\begin{equation}
    \tilde{V}_{IJKL} = \phi^*_{iI} \phi^*_{iJ} \phi_{jK} \phi_{jL} V_{ij}
\end{equation}
where we use capital letters to label the indices of the set of target orbitals with known two-electron integral, and lower case to label the basis functions. Here 
$V_{ij}$ is the two electron interaction matrix of the hybrid basis functions determined by MWG. 
Suppose we know $V_{IJKL}$, the exact values of the integrals. 
We can add a correction $\delta V_{ij}$ to $V_{ij}$, while keeping the relative size of the correction small, to try to fix the errors in the diagonal approximation. A very natural form of correction is to minimize
\begin{equation}
         \sum_{i,j} \delta V_{ij}^2 
         \label{Eqn:correctVnn1}
\end{equation}
over $\delta V$, subject to
\begin{equation}
        \sum_{i,j} A^*_{i,IJ} A_{j,KL} (V_{ij}+\delta V_{ij})   = V_{IJKL}
        \label{Eqn:correctVnn2}
\end{equation}
where $A_{i,IJ} = \phi_{iI} \phi_{iJ}$.
By treating $IJ$ as a composite index (as well as $KL$), and introducing a Lagrange multiplier, the problem becomes minimizing the Lagrangian written in matrix form, 
\begin{equation}
    \mathcal{L} = \rvert\rvert A^{\dagger} \delta V A - B \lvert\lvert^2 + 
    \lambda \rvert\rvert \delta V \lvert\lvert^2 ,
\end{equation}
where $B_{IJ,KL} = V_{IJKL} - \sum_{i,j} A^*_{i,IJ} A_{j,KL} V_{ij}$, and the matrix norm is $\rvert\rvert A \lvert\lvert^2 = \mathrm{tr} (A^{\dagger} A)$. The minimizer of the Lagrangian satisfies 
\begin{equation}
    \lambda \, \delta V + AA^{\dagger} \delta V AA^{\dagger} 
    = ABA^{\dagger},
\end{equation}
which we can solve by a singular value decomposition  $A = U \Sigma W^{\dagger}$.
Denote $\delta \tilde{V} \equiv U^{\dagger} \delta V U $ and
$\tilde{B} = W^{\dagger} B W$, then
\begin{equation}
    \lambda \, \delta \tilde{V} + \Sigma^2 \delta \tilde{V} \Sigma^2 
    = \Sigma \tilde{B} \Sigma,
\end{equation}
in other words,
\begin{equation}
    \delta \tilde{V}_{\mu\nu} =
    \frac{\tilde{B}_{\mu\nu}\sigma_{\mu}\sigma_{\nu}}
    {\lambda + \sigma^2_{\mu}\sigma^2_{\nu}},
\end{equation}
where $\sigma_{\mu}$ denotes an individual singular value.
With $\delta \tilde{V}$ solved, $\delta V$ can be obtained through $\delta V = U \delta \tilde{V} U^{\dagger}$. The Lagrange multiplier controls the (square of) errors in fitting $V_{IJKL}$, and it is set to be around $10^{-18}$ in our applications. 

Usually, there are too many Gaussians in a full standard basis to apply this correction. It may
be possible to fit the $V_{IJKL}$ exactly in some cases, but only at the cost of large changes in the $V_{ij}$.  This can make the Hamiltonian much worse for the orbitals that are not explicitly corrected. We find that requiring the maximum relative changes in $V_{ij}$ to be no more than about 10\% 
does not lead to overfitting. To reduce the number of target orbitals, we contract the Gaussians.
A simple approach that we have adopted is to perform a single-atom Hartree-Fock calculation for each type of atom using a highly-accurate Gaussian basis (e.g. cc-pV6Z). We then contract to the occupied
atomic orbitals. This gives us a small number of highly-relevant orbitals to target for correction.
We call this general approach the Exact Gaussian Orbital Integral correction (EGOI), and in the
calculations performed here we have only used the atomic-HF version of this correction. However, this is clearly an area in which further improvements could be made. 

We have also utilized a simpler way of applying a correction in the case of correcting the hydrogen-like 1s orbital,
which, of course, is very localized near the nucleus.
In this case we choose the target orbital $\phi$  as the lowest eigenvector of the one-body Hamiltonian where we only include the potential of the given nucleus. Thus $\phi$ (with basis coefficients $\phi_j$) is our best approximation to the exact 1s orbital, which is quite good because of the Gaussian basis functions. In this correction, for each atom we modify only the single two electron matrix element $V_{ii}$ to make the lattice give this exact result. 
Assume that the $i$-th gausslet is located at a certain nucleus (or at least $i$ is the closest gausslet). 
The Hamiltonian gives
$\bra{1s\,1s}V\ket{1s\,1s} = \sum_{j,k} \phi_j^2 V_{jk} \phi_k^2$.  
We then correct this value by modifying the single matrix element $V_{ii}$ as
\begin{equation}
    \delta V_{ii} = \frac{1}{\phi_i^4} (\frac{5}{8}Z - \sum_{j,k}\phi_j^2 V_{jk} \phi_k^2).
\end{equation}
We call this approach the Exact Slater Orbital Integral correction (ESOI).
The correction can be done for each atom separately.  In principle, if core levels overlap, the corrections for different atoms may influence each other, but this effect is small and another pass through applying corrections reduces it much further.  

The EGOI correction can be carried out using Eqs. (\ref{Eqn:correctVnn1}) and (\ref{Eqn:correctVnn2}).
Since both ESOI and EGOI correct the atomic $V_{ij}$, one should not use both.  In the case of a hydrogen atom, these two corrections are almost identical, and we use ESOI, while for other atoms we use EGOI.

{\it \underbar{Stationary Fock correction}} 
Suppose the occupied HF orbitals $\phi_I$ are known and the hybrid gausslet-Gaussian basis can well represent them. 
The occupied HF orbitals can be extended into an orthonormal basis of the space spanned by the hybrid gausslet-Gaussian basis.
The Fock matrix can be constructed and cast into this MO basis. 
But, due to the fact that the hybrid basis is not exactly complete,
and that the diagonal approximation of two-electron integrals also introduces small error, the matrix elements of the occupied-virtual block of this Fock matrix will slightly deviate from zero.

The occupied-virtual block of the Fock matrix can be zeroed out by adding a correction term to the one-body Hamiltonian. 
Let $\phi_{iI}$ be the coefficient of $I$-th HF orbital on the $i$-th basis. The correction term can be calculated as
\begin{equation}
    \Delta h_{ij} = - \sum_{\substack{ i \in \text{occ} \\ A \in \text{vir} }} \phi^*_{iI} \phi_{jA} F_{IA}
\end{equation}
where $I$ indexes the occupied HF orbitals, while $A$ indexes the virtual ones.

We applied the Stationary Fock correction at the atomic level, because highly accurate HF calculations from atoms are readily available. And, the atomic HF orbitals can be pre-calculated and stored for repeated use.

\begin{figure}[t]
    \includegraphics[width=0.96\columnwidth]{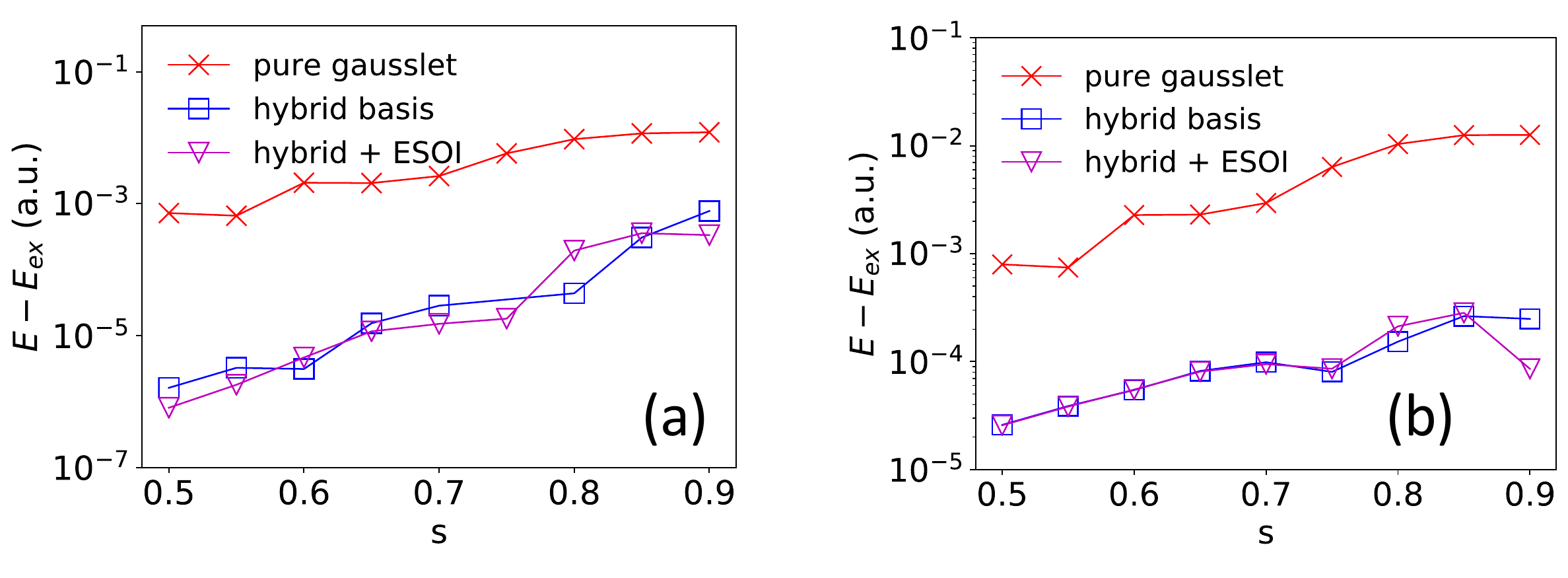}
    \caption{
    Assessing the effects of the ESOI on H$_2$ molecule at R=2.0 bohr.
    Errors of (a) Hartree-Fock energies and (b) exact diagonalization are shown, versus the spacing parameter $s$ (which equals $c$ for hydrogen atom here).
    ``Pure gausslet" denotes the uncorrected pure gausslets basis set.
    ``Hybrid basis" means the Gaussians are added to the basis set,
    and MWG approximation is applied to their two-electron integrals.
    ``Hybrid + ESOI" means the onsite two-electron integrals are corrected to reproduce the exact onsite two-electron integral of 1S orbital.
    The cc-pV6Z Gaussians, with up to P orbitals, are added.
\label{Fig:H2_hf+fci}}
\end{figure}

To assess the effects of various corrections on the Hamiltonian,
the HF and exact energies of H$_2$ molecule are shown in Fig. \ref{Fig:H2_hf+fci}. 
There are some kinks on the energy curves,
which results from gausslets passing through nuclei while the spacing parameter $s$ is decreased.
For all variants of gausslets, both the HF and FCI energies converge nicely to the CBS limit as the scale of gausslets decreases. 
And noticeably, adding in the Gaussian orbitals significantly accelerates the convergence. 
The error in the HF energy is brought down from $10^{-3}$ to around $10^{-6}$ Hartree at $s=0.5$, 
while the error of the exact energy is reduced from $10^{-3}$ to around $10^{-5}$ Hartree at $s=0.5$.
Without the Gaussians, a small spacing parameter, and thus a large number of gausslets, is needed to describe the cusps.
The ESOI correction gives slightly better HF energies at most of the scales, but overall its effect is very small.


\begin{figure}[t]
    \includegraphics[width=0.96\columnwidth]{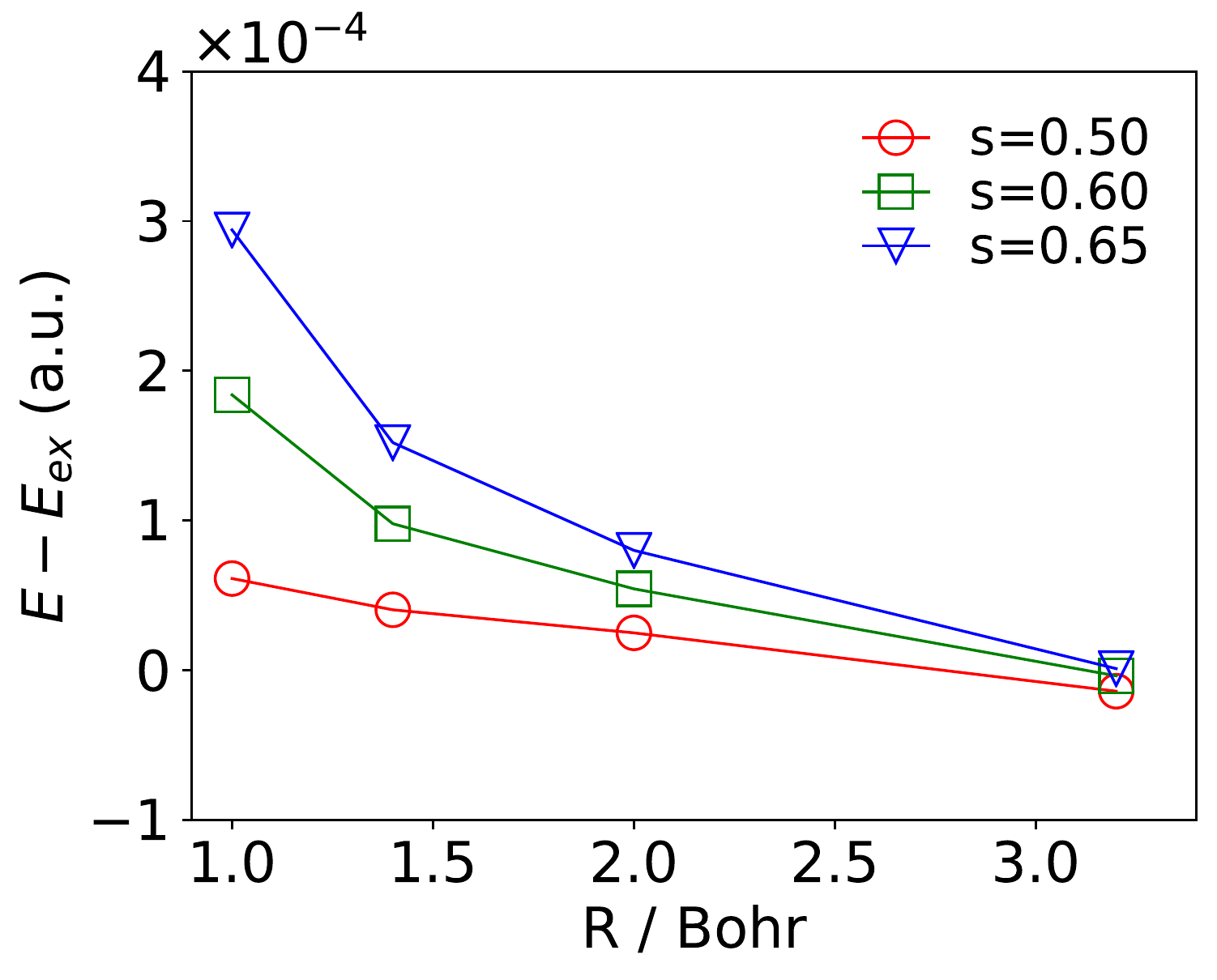}
    \caption{
    Assessing the quality of hybrid basis set at different bond lengths.
    Errors of H$_2$ exact ground-state energies at different bond lengths.
    The coresize and scale parameter are set to be the same.
    Gaussians orbitals added are from cc-pV6Z basis set, with up to P orbitals.
    The ESOI correction is applied.
\label{Fig:H2_fci_R}}
\end{figure}

The performance of gausslet basis sets at different bond lengths are shown in Fig. \ref{Fig:H2_fci_R}.
As the spacing parameter $s$ decreases, the basis set is systematically improved across different bond lengths,
and the error curve gets closer and closer to a horizontal flat line. 
Also shown in the plot, small bond lengths are the more difficult cases, 
giving rise to larger errors.
This can probably be explained by the fact that
wavefunctions at small bond lengths oscillate more intensively,
which entails higher-frequency gausslets and thus smaller scales.


\begin{figure}[t]
    \includegraphics[width=\columnwidth]{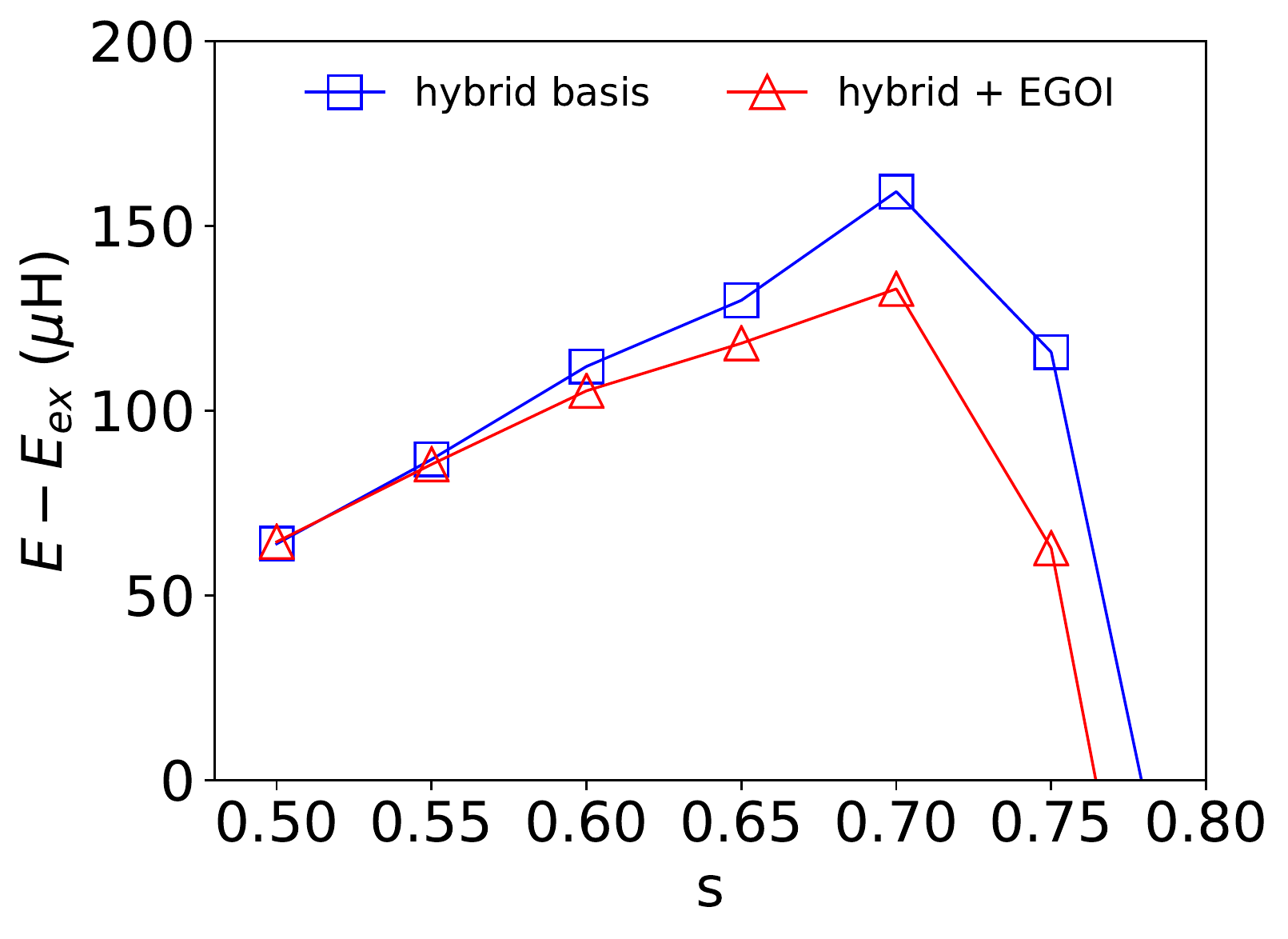}
    \caption{
    Assessing the effects of Hamiltonian corrections on exact two-electron energies.
    The plot shows errors of He atom exact ground-state energies versus the spacing parameters $s$ of the gausslets. 
    The core size parameter is set to be $c=0.5s$.
    ``Hybrid basis" denotes the hybrid gausslet basis set with Gaussians added in,
    and with MWG approximation applied to their two-electron integrals.
    The cc-pV6Z Gaussians, with up to P orbitals, are added.
    In the curve denoted by ``Hybrid + EGOI", both EGOI and stationary Fock corrections are applied.
    In EGOI, the two-electron integrals are corrected so as to reproduce the two-electron integrals of a known set of HF orbitals.
    In ``stationary Fock", the one-electron integrals are corrected so that the known atomic HF orbitals remain stationary.
\label{Fig:He_fci_scale}}
\end{figure}


Similar tests for the helium atom are shown in Fig. \ref{Fig:He_fci_scale} where we can see the effect of the EGOI correction.
Similar to the case of the H$_2$ molecule, the most significant correction is adding the Gaussians to the basis, 
which brings the error of exact energy from around $10^{-3}$ (not shown) down to $10^{-4}$ Hartree.
The result shown shows the EGOI with the stationary Fock correction also put in, but the stationary Fock part does not affect the results visibly in this system (not shown). 
Below, we have turned on the stationary Fock correction whenever we use EGOI, but it could also be omitted. 
The EGOI  itself improves the energy a little bit at larger scales.  The most significant errors, causing the slow convergence of both curves in Fig. \ref{Fig:He_fci_scale} with $s$, are not associated with the diagonal approximation.  Instead, they are due to the electron-electron cusp, and are discussed in the next section.

\begin{figure}[t]
    \includegraphics[width=\columnwidth]{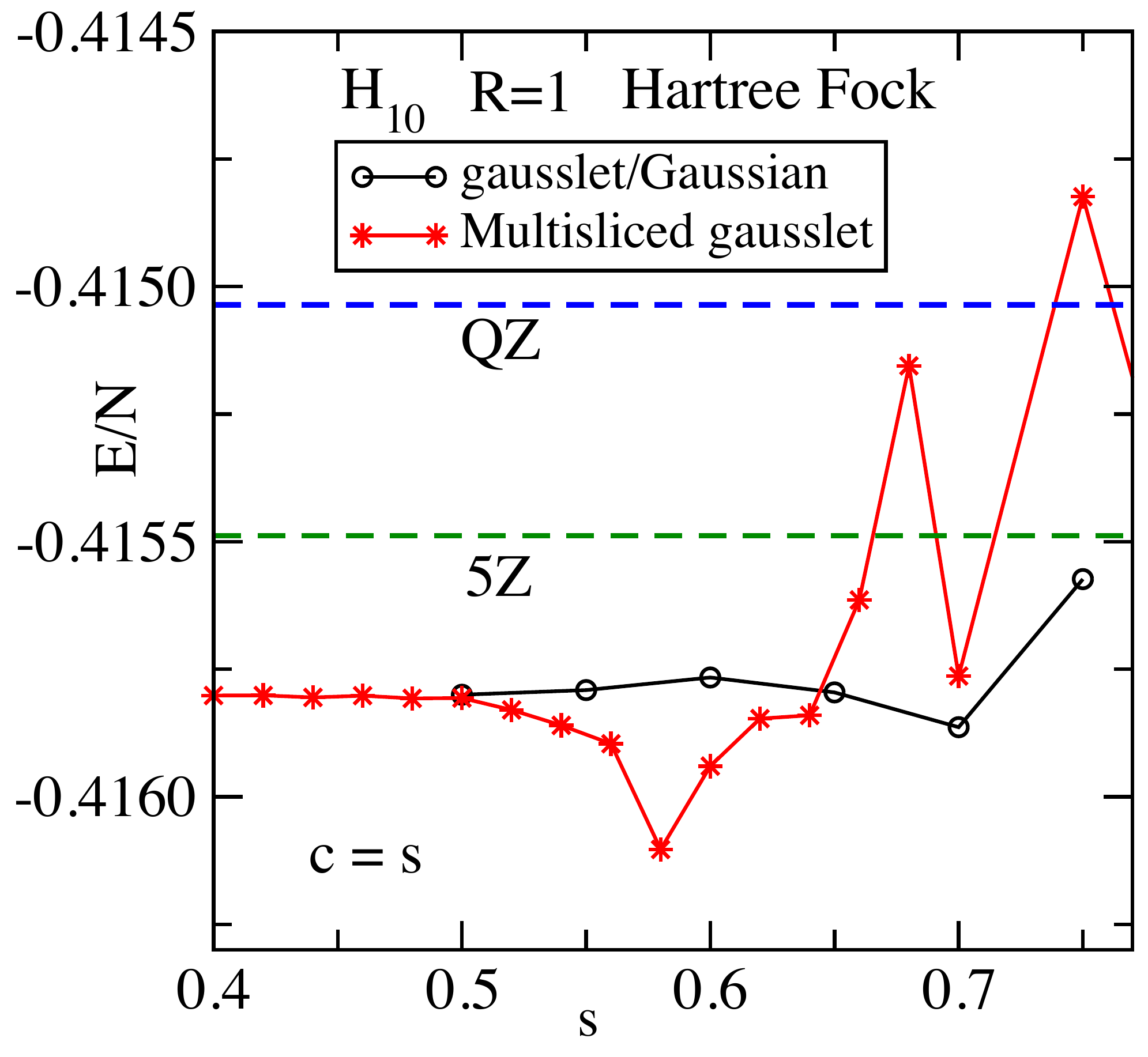}
    \caption{
    Hartree Fock results for $H_{10}$ at $R=1$, comparing pure Gaussian bases, gausslet/Gaussian bases, and multisliced gausslet bases of \citet{White19}.   The multisliced bases did not have Gaussians, instead using core corrections only to accelerate convergence.  The gausslet/Gaussian bases show more rapid convergence with s, where we take $c=s$. 
\label{Fig:H10_hf}}
\end{figure}

To verify that our conclusions about constructing the Hamiltonian are not somehow specific to two-electron systems, we have used them to compute HF energies for the linear equal-spaced chain H$_{10}$ at a spacing R=1.0 bohr, shown in Fig. \ref{Fig:H10_hf}. (Excellent behavior at the full-CI level (with DMRG, without the Gaussians) in H$_{10}$ was demonstrated  in Ref. \onlinecite{White19}.)  This distance is short enough that the pure Gaussian bases (cc-pVQZ and cc-pv5Z) are less accurate than at larger separation, and it is difficult using them to get high accuracy, say 0.1 mH/atom. With the gausslet bases, it is straightforward to converge to well below 0.1 mH/atom. We do not have a superior result with some other approach to determine the exact size of the final errors, but it appears to be around the $10^{-5}$ Hartree level. Note that Hartree Fock does not require resolution of the electron-electron cusps, so the accuracy here is much higher than in the full-CI H$_2$ or He results. The convergence with $s$ is more rapid with the gausslet/Gaussian bases than the multisliced bases, but for a fixed $s$, the multislicing results in  fewer functions.  In future work we will show how to reduce the number of basis functions in
the gausslet/Gaussian bases below that of the multisliced bases, while maintaining high accuracy.

\section{Corrections for the Electron-Electron Cusp}

In this section we present a very simple correction for the basis set incompleteness energy error due to
the electron-electron cusp. While the correction is simple, with only two universal parameters, it is also
important to consider the ideas behind it, and how we came up with this correction.  Our original
effort was to correct the Hamiltonian terms for the cusp, but in the end we found a post-calculation
correction to the energy works very well.

The electron-electron cusp occurs at every point in space where two electrons meet, so it cannot
be cured by adding a few specialized basis functions.  However, the local nature of the gausslets
makes it possible to devise corrections for the part of the cusp below the resolution of the gausslets.
A significant part of the cusp occurs when two electrons occupy the same gausslet.
In contrast, when two electrons are on gausslets which are not close, the cusp is irrelevant.
A conventional
two electron integral for the onsite term assumes no correlation within the gausslet.
One could imagine replacing the integral with either a modified pseudo electron-electron interaction,\cite{Fahy}
or perhaps with a Jastrow correlation factor inserted.
Thus one might correct for the cusp by adjusting the onsite terms $V_{ii}$.
The onsite Hamiltonian contributes to the energy
\begin{equation}
E_{\rm onsite} = \sum_i V_{ii} d_i
\end{equation}
where $d_i = \langle n_{i\uparrow} n_{i\downarrow} \rangle$ is the double occupancy. Note that in the limit
of a complete gausslet basis, both $d_i$ and $D=\sum_i d_i$ tend to zero, and these quantities can
be used as measures of the (in)completeness.

It is simplest to consider correcting only $V_{ii}$ with a correction $\delta V_{ii}$, but adjusting the
correction to take into account both the on-site and nearby-site cusp errors.  This approach is conceptually
similar to using a local exchange-correlation potential in density functional theory, but note that
the correction and any errors in it will vanish as the basis is made more complete. The key question is,
can we make a sensible ansatz for $\delta V_{ii}$?  Perhaps we can find a function $\delta V(V_{ii},d_i,n_i)$ (where
$n_i$ is the density on site $i$) which provides a good correction.

It is useful to consider the scaling of the wavefunction and energy of two electron atomic ions, which are
readily solved to high accuracy with the same techniques as for a helium atom.  As $Z$ increases, the
radius of the 1S natural orbital falls as $1/Z$, but the total energy increases roughly as $Z^2$. We could form
a good gausslet basis for the ion by taking the He atom basis and scaling the distances and sizes of all functions by
$2/Z$.  A surprising but long-known observation about the two electron
ions is that their correlation energies are almost constant in $Z$\cite{DHCUF91,RFHS87}. As $Z$ increases, the increase in the
total energy is offset by the ions becoming more weakly correlated.

The cusp energy errors can be considered as a part of the correlation energy. Thus one would expect the local correction
terms  $\delta V_{ii} d_i$ to be approximately constant as $Z$ is changed. As $Z$ changes, the $V_{ii}$ scale as $Z$,
but $d_i$ and $n_i$ remain roughly constant in $Z$ (using our scaled He basis).
This tells us that our correction should be constant in $V_{ii}$, depending as
$\delta V(d_i,n_i)$.

Since the cusp is associated with two electrons at the same location, it is much more naturally tied to $d_i$ than to
$n_i$, so
we consider only $\delta V(d_i)$.
Now consider a fixed atom, say He, and change $s$, which $c=s/Z$. We observe that the total double occupancy $D$
and the error in the total energy $\Delta E$ scale as powers of $s$
\begin{equation}
    \Delta E \sim s^\gamma
\end{equation}
\begin{equation}
    D \sim s^\beta
\end{equation}
with $\gamma \sim 3.4$ and $\beta \sim 5$.  Thus, $\Delta E \sim D^{\gamma/\beta}$.
Considering larger systems, say many separated He atoms, we note that a correction that depended on a power of $D$ would not be size consistent.
However, connecting back to a correction to the Hamiltonian, where size consistency is easy to obtain, and assuming the $d_i$ behave similarly
to $D$, we make the following ansatz for the electron-electron cusp basis set error
\begin{equation}
    \Delta E = e_0 \sum_i d_i^\alpha
\end{equation}
This correction can be use directly, or used to as a correction to the onsite interactions $V_{ii}$,
$\delta V_{ii} = e_0 d_i^{\alpha-1}$.  However, since the correction requires $d_i$, which one knows only
after one solves the system, this form requires a self-consistent calculation.  Thus it is simpler to
attempt only an energy correction $\Delta E$.

We considered a data set of about a dozen different two-electron systems and attempted to find universal values for $e_0$ and $\alpha$
which would work for all.
The systems included the He atom with different $s$ (with $c=s/2$), the Be atom with different $s$ (with $c=s/4$),
and H$_2$ at $R=2$ for a few different $s=c$.  It was not difficult to find values that did well on all systems.
We find
\begin{equation}
e_0 = -0.005078
\label{Eqn:cusp_e0}
\end{equation}
\begin{equation}
\alpha = 0.79
\label{Eqn:cusp_alpha}
\end{equation}

\begin{figure}[t]
    \includegraphics[width=\columnwidth]{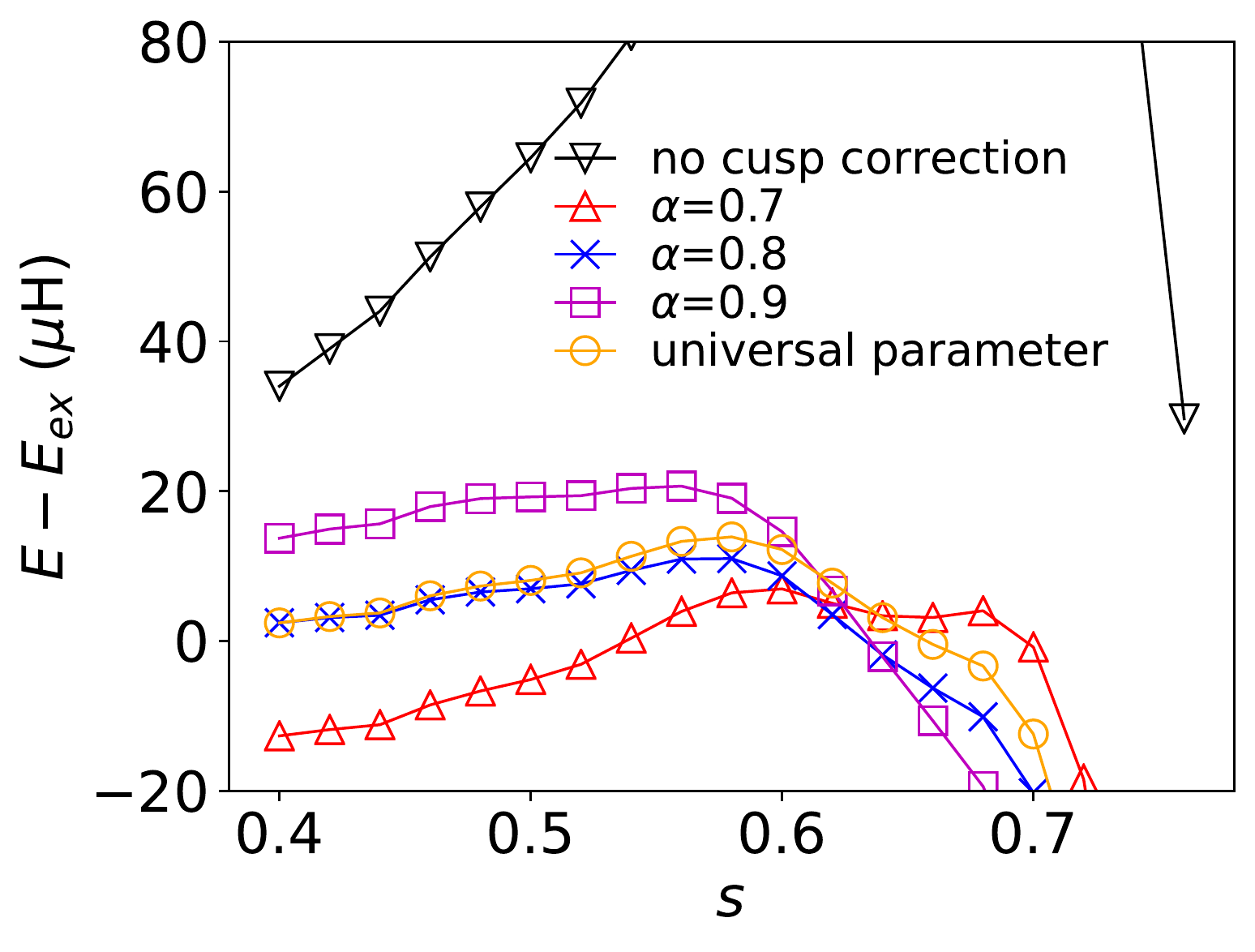}
    \caption{
    Scanning through different exponents $\alpha$ used in the two-electron cusp correction. The plot shows the exact energy of He atom versus the spacing parameter $s$, where the core size parameter is set to be $c=0.5s$. The hybrid basis set uses cc-pV6Z Gaussian with up to P orbitals.
    For the curves labeled by $\alpha$, the coefficient parameter $e_0$ is optimized for the He atom for $s$ between $0.4$ and $0.7$.
    The curve labeled by ``universal parameter" uses parameters optimized for both He atom, Be atom and H$_2$ molecules at several bond lengths, which is given by Eqn. \ref{Eqn:cusp_e0}, \ref{Eqn:cusp_alpha}.
\label{Fig:He_cusp_corr}}
\end{figure}
\begin{figure}[t]
    \includegraphics[width=\columnwidth]{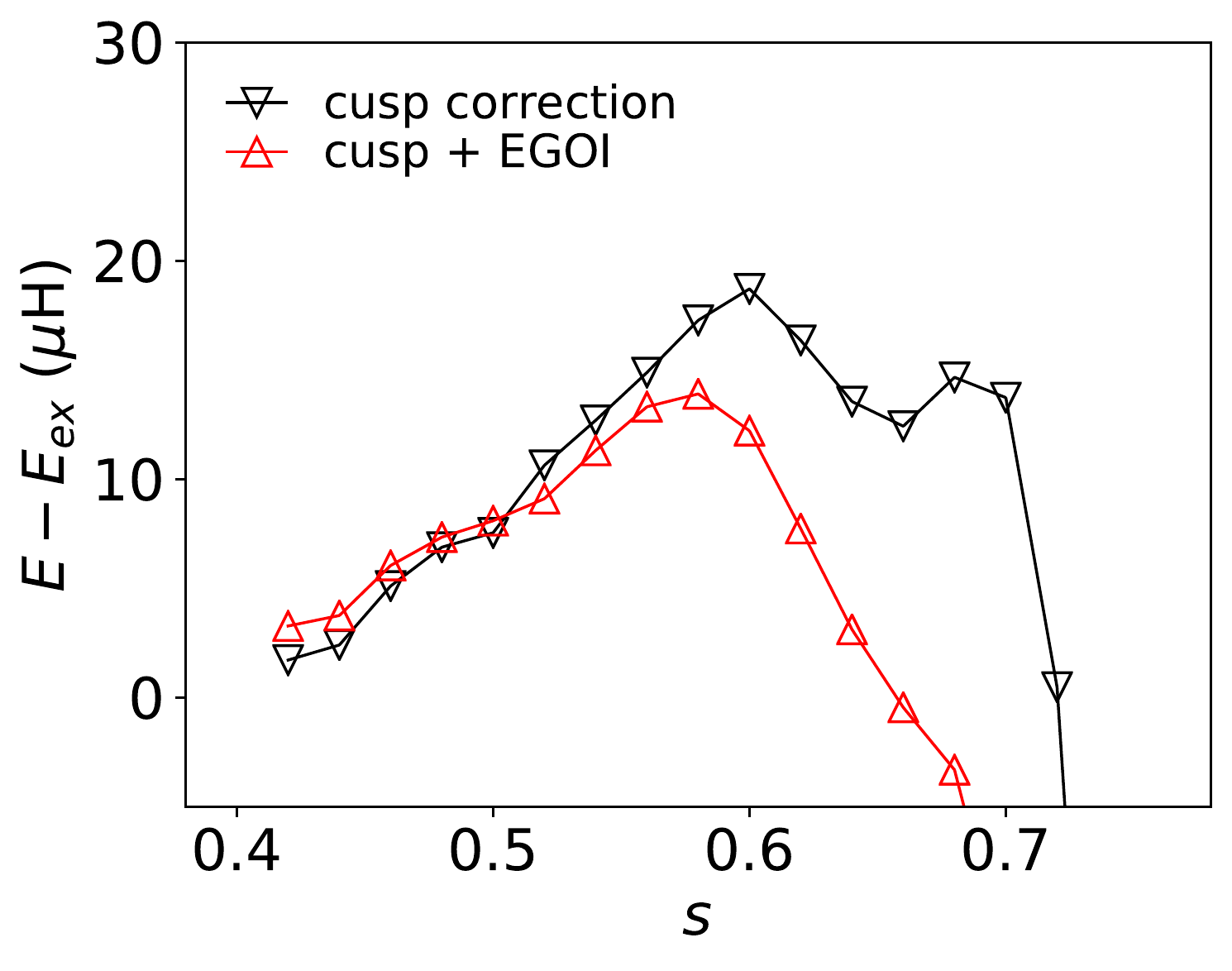}
    \caption{
    Two electron cusp correction applied to He atom at different spacing,
    where the core size parameter is set to be $c=0.5s$.
    The hybrid basis set uses cc-pV6Z Gaussian with up to P orbitals.
    The universal parameter for cusp correction is used.
    The curve labeled by ``cusp + EGOI" also put in the EGOI and stationary Fock corrections.
\label{Fig:He_cusp_corr_EGOI}}
\end{figure}
\begin{figure}[t]
    \includegraphics[width=\columnwidth]{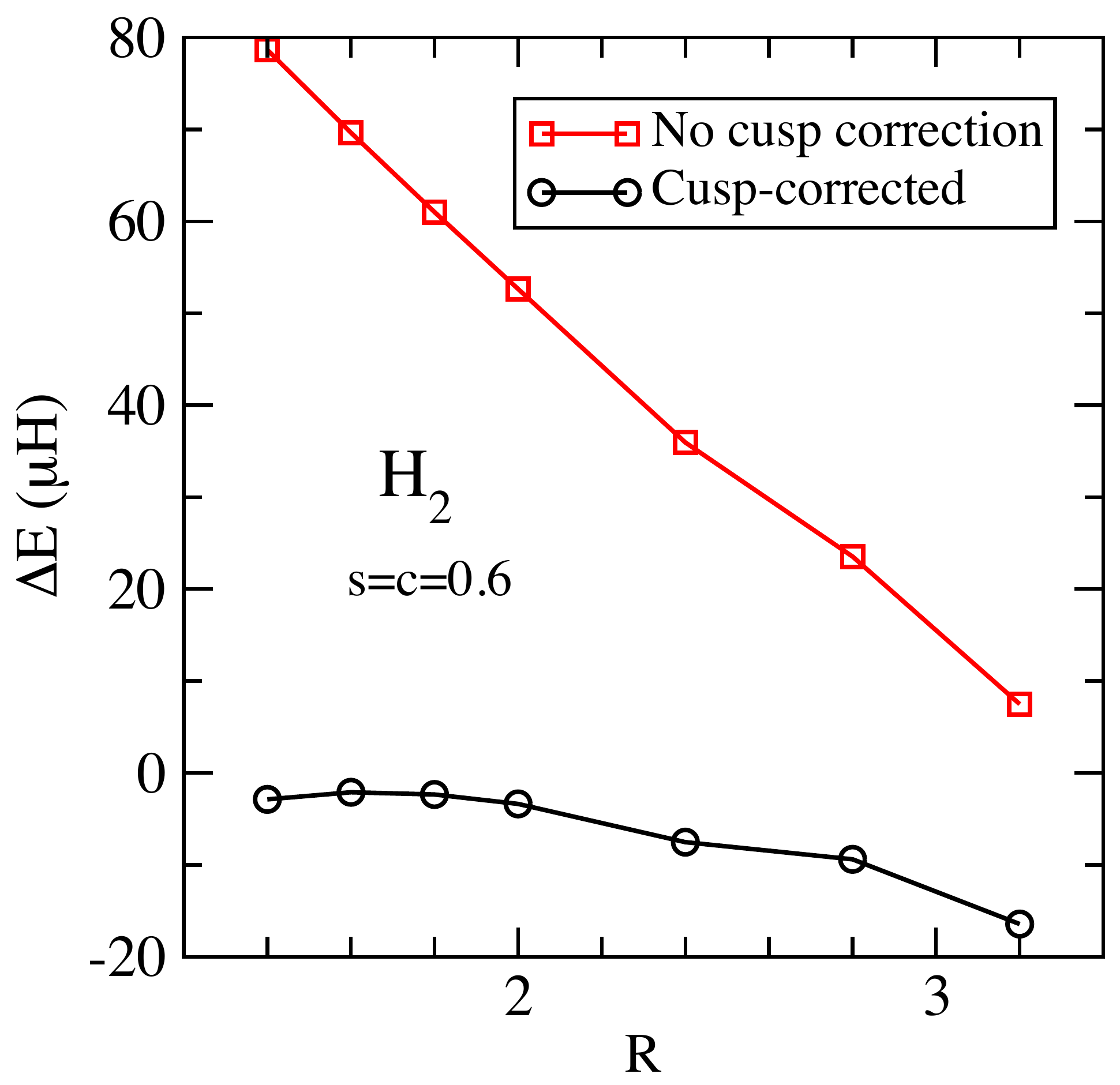}
    \caption{
    Two electron cusp correction applied to H$_2$ molecule at different bond lengths. 
    The spacing and core size parameters are set to be $s=c=0.6$.
    The hybrid basis set uses cc-pV6Z Gaussian with up to P orbitals.
\label{Fig:H2_cusp_corr}}
\end{figure}
To illustrate the process of optimizing the exponent $\alpha$, Fig. \ref{Fig:He_cusp_corr} show fits with a few different $\alpha$'s in the case of the He atom. As shown in the figure, $\alpha$ around $0.79$ gives the overall best result for most of the scales, especially in the asymptotic small $s$ region. 
Also note that the parameter $e_0$ optimized for He and the universal $e_0$ given by Eqn. \ref{Eqn:cusp_e0} gives similar results, which is a good indication of the generality of the two-electron cusp correction.

The parameter given by Eqn. \ref{Eqn:cusp_e0} and \ref{Eqn:cusp_alpha} works very well for two electron systems such as H$_2$ molecules and the He atom, as shown in Fig. \ref{Fig:He_cusp_corr_EGOI} and \ref{Fig:H2_cusp_corr}. 
The two-electron cusp correction brings the error of ground-state energies down to the micro Hartree level, consistently for different bond lengths of the H$_2$ molecule, and for different $s$ in the case of He atom. With the cusp correction accounted for, we can revisit the EGOI correction. While the EGOI and stationary Fock corrections only have a  small effect on the total energies, as shown in Fig. \ref{Fig:He_cusp_corr_EGOI}, at larger $s$ the correction in this case is comparable in size to the errors not accounted for by the cusp, and in the right direction. Thus, currently, the EGOI correction is not especially useful, but it is worth considering in very high accuracy calculations.

\section{Conclusion}
The locality, systematic completeness, orthogonality, variable resolution, and special moment properties of gausslet bases gives them key advantanges for electronic structure calculations, particularly for DMRG and perhaps tensor networks.  In particular, the moment properties give rise to a 
 diagonal approximation, 
reducing the number  of two-electron integrals from $N^4$ down to $N^2$. 
This in turn reduces the computational scaling of electronic structure methods, such as Hartree-Fock and DMRG.  In this paper we have shown how one can add
a limited number of Gaussians to a gausslet basis, to accelerate convergence in
the core regions. We have shown how one can maintain the diagonal $N^2$ form of the interactions in the presence of the added Gaussians.   We demonstrate good and fast approximations for evaluating the integrals involving the orthogonalized Gaussians. These approximations and the diagonal aproximation itself work well
because most of the occupancy occurs in the gausslet part of the basis, where
the diagonal approximation is on firm ground. 

We have also shown that one can further correct errors in the approximate
interactions by requiring known integrals of orbitals involving the whole basis
be reproduced exactly. The ideal form of this would be to make all integrals among a set of leading natural orbitals be reproduced exactly, but as a surrogate for this one can correct for approximate Hartree Fock orbitals or simply some of the Gaussians that have been added to the basis.  A limitation is that the diagonal form only allows a limited number of orbitals to be corrected for.  The number of
corrections we can put in ends up not having a very big effect of the results, since our initial approximations performed well. 

We have also tried a different type of correction, where we correct for the
incomplete resolution of the electron-electron cusp.  This type of correction has
been developed before for standard  basis, and these corrections are related to
F12 and R12 methods.\cite{KBFVE12,HKKT12}  However, the
local nature of the gausslets makes it easy to devise a simple correction.
In this case a major part of the cusp error comes from two electrons being on the same gausslet. Our correction is thus tied to the double occupancy of each basis function. In a number of two electron systems we observed that the cusp error behaves nonanalytically with the double occupancy, and we propose a simple two-parameter form for the energy correction as a function of the double occupancies.  We find a single pair of parameters which work well for a number of two-electron systems, including
the He atom, the Be atom, and H$_2$ molecules at different bond lengths, in each case with different grid sizes. 
The universal cusp correction works very well for these two-electron systems,  bringing the full-CI complete-basis energy errors down to the level of microHartrees.

While many-electron wavefunctions are exponentially more complicated than two electron wavefunctions, the Hamiltonians themselves are the same.  Here, since we are working to construct Hamiltonians rather than wavefunction approximations, we believe that the two-electron calculations we mostly use here as test cases show us how to get good Hamiltonians for use in the many particle case. In the case of the cusp correction, we have not accounted for the higher order singularities when two electrons with the same spin approach one another.  This would be an example of a difference with the many-electron case, although one might study two-electron systems in a triplet state. But certainly the biggest cusp basis-incompleteness error comes from opposite spins, so we believe our cusp correction will be helpful in the many-electron case. 

The hybrid basis construction can be followed by subsequent decimation procedures to  reduce the size of the basis. This reduction will greatly improve the ability of these bases to be used for DMRG and other techniques in larger systems, and it will be important to revisit our corrections and techniques in that context.

\begin{acknowledgments}
We thank Garnet Chan, Randy Sawaya, Kieron Burke, Ryan Pederson, and Miles Stoudenmire for helpful discussions.  This work is funded by NSF through Grant DMR-1812558 (SRW),  and through  U.S. Department of Energy,
Office of Science, Basic Energy Sciences under award DE-SC0008696 (YQ), and by the Simons Foundation through the Many-Electron Collaboration (YQ). 
\end{acknowledgments}

\bibliography{ref.bib}

\appendix
\section{Wavelet Approach to Approximate Two-electron Integrals of Residual Gaussians}
1D gausslets are related to scaling functions of ternary wavelet transformations. We can apply one additional wavelet transform on an array of gausslets to construct a set of ``scaling functions'' and ``wavelets''.  Since the
gausslets are not actual scaling functions, the functions we obtain are also not true scaling functions and wavelets, but for most practical purposes they act very much like them. True scaling functions are fixed points of a wavelet transform.  As part of a multiscale representation, true wavelets at different levels are scaled and shifted copies of a single function.  Since we do not use a multiscale analysis, this property is not relevant. What is relevant is that an additional wavelet transform or two applied to a gausslet basis retains all the desirable completeness, orthonormality, and moment properties of the gausslets. 
After a WT, the transformed functions can still be represented as a sum over a uniform array of Gaussians, but the spacing of the underlying Gaussians is now $1/9$ the spacing of the functions instead of $1/3$. 

For simplicity, we will refer to our transformed gausslets as scaling functions and wavelets. For our ternary WT transformations,
each 1D scaling function comes with two adjacent wavelets. 
Since 3D basis functions are direct products of 1D ones, each 3D scaling function is 
associated with $3^3-1=26$ wavelets.
The 3D wavelets have the form $W(x)G(y)G(z)$, $W(x)W(y)G(z)$ or $W(x)W(y)W(z)$, plus permutations,
where $G, W$ denote 1D scaling function and wavelet, respectively. To create this scaling/wavelet basis, we reduce the parameter $s$ to $s/3$ (with $c$ held fixed) and construct 1D gausslet bases, and then apply 1D WTs to them to get 1D scaling functions and wavelets as scaling parameter $s$. We then can construct the appropriate 3D product functions out of the 1D functions. 
The resulting basis has the completeness properties of the $s/3, c$ basis it came from, but we think of it as being a basis at scale $s, c$, but with extra wavelets for additional completeness.  Not all of the extra wavelets need to be included---in fact, we should only include them where additional resolution is needed.

While one could use wavelets as a substitute for the Gaussians we add to the basis, they are not as efficient--many more would be needed to capture the core features of a single Gaussian.  However, we can use them instead as an alternative way of constructing the diagonal approximation of the residual Gaussians.  Specifically, the residual Gaussians can be approximated by a linear combination of wavelets, i.e.,
\begin{equation}
    g_I = c_{iI} W_i + \tilde{g}_I,
\end{equation}
where $g_I$ denotes a residual Gaussian, $W_i$ is a 3D wavelet, and $\tilde{g}_I$ is
the leftover piece of the Gaussian that is orthogonal to both the gausslets and wavelets (of the current scale).
The leftover piece $\tilde{g}$ is not zero because
the wavelets are not complete in the orthogonal space of gausslets, 
and wavelets of smaller scales are needed to achieve exact completeness. 
However, the wavelets should capture the major parts of the residual Gaussians, 
and $\tilde{g}$ is small enough to be neglected.

Wavelets are localized in real space just as gausslets.  However, they do not have the delta-function property.  Nevertheless, given their small occupancy, it is reasonable to apply a
diagonal approximation of the two-electron integrals applies to them. In this case, we take as the diagonal approximation a simple truncation of all the off diagonal terms in a standard $V_{ijkl}$. 
Thus, we only consider two-electron integrals that are of the form $V_{iijj}$.
This gives for any term involving residual Gaussians the diagonal form
then
\begin{equation}
    V_{IIJJ} \approx V_{www'w'} c^{*2}_{wI} c^2_{w'J},
    \label{Eqn:contractWave}
\end{equation}
where $I,J$ index the residual Gaussians, $w,w'$ label the wavelets, 
and $c_{wI}$ is the expansion coefficient of the $I$-th residual Gaussian on the $w$-th wavelet.
Since the wavelet two-electron integrals and expansion coefficients are known, the equation above gives an approximation to the diagonal two-electron integrals of the residual Gaussians.

\end{document}